\documentclass[sigconf, xcdraw, table, screen]{acmart}

\renewcommand\footnotetextcopyrightpermission[1]{} 

\usepackage{listings}
\usepackage{colortbl}
\usepackage{algorithm}
\usepackage{algpseudocode}

\definecolor{codegreen}{rgb}{0,0.6,0}
\definecolor{codegray}{rgb}{0.5,0.5,0.5}
\definecolor{codepurple}{rgb}{0.58,0,0.82}
\definecolor{backcolour}{rgb}{0.95,0.95,0.92}

\newcommand{\etal}{et al.\ }

\lstdefinestyle{mystyle}{
    commentstyle=\color{codegreen},
    keywordstyle=\color{magenta},
    numberstyle=\tiny\color{codegray},
    stringstyle=\color{codepurple},
    basicstyle=\ttfamily\footnotesize,
    breakatwhitespace=false,         
    breaklines=true,                 
    captionpos=b,                    
    keepspaces=true,                 
    numbers=left,                    
    numbersep=5pt,                  
    showspaces=false,                
    showstringspaces=false,
    showtabs=false,                  
    tabsize=2
}

\AtBeginDocument{%
  \providecommand\BibTeX{{%
    \normalfont B\kern-0.5em{\scshape i\kern-0.25em b}\kern-0.8em\TeX}}}

\copyrightyear{2023}
\acmYear{2023}
\setcopyright{rightsretained}
\acmConference[UIST '23]{The 36th Annual ACM Symposium on User Interface Software and Technology}{October 29-November 1, 2023}{San Francisco, CA, USA}
\acmBooktitle{The 36th Annual ACM Symposium on User Interface Software and Technology (UIST '23), October 29-November 1, 2023, San Francisco, CA, USA}
\acmDOI{10.1145/3586183.3606723}
\acmISBN{979-8-4007-0132-0/23/10}

\newif\ifsubmit
\submitfalse
\ifsubmit
    \newcommand{\stefanie}[1]{}
    \newcommand{\mackenzie}[1]{}
    \newcommand{\faraz}[1]{}
    \newcommand{\ahmed}[1]{}
    \newcommand{\changes}[1]{}

\else
    \newcommand{\stefanie}[1]{{\leavevmode\color[rgb]{1.0, 0.0, 0.5}{#1}}}
    \newcommand{\mackenzie}[1]{{\leavevmode\color[rgb]{1.0, 0.6, 0.0}{#1}}}
    \newcommand{\faraz}[1]{{\leavevmode\color[rgb]{0.5, 0, 1.0}{#1}}}
    \newcommand{\changes}[1]{{\leavevmode\color[rgb]{0.0, 0.0, 0.0}{#1}}}
    \newcommand{\ahmed}[1]{{\leavevmode\color[rgb]{0, 0.7, 1.0}{#1}}}
\fi

\usepackage{soul}

\usepackage{booktabs}
\usepackage{multirow}
\usepackage{graphicx}
\usepackage{microtype}

\begin{document}




\title{Style2Fab: Functionality-Aware Segmentation for Fabricating Personalized 3D Models with Generative AI}



\author{Faraz Faruqi}
\email{ffaruqi@mit.edu}
\orcid{0000-0002-1691-2093}
\affiliation{%
  \institution{MIT CSAIL}
  \city{Cambridge}
  \state{MA}
  \country{USA}
}

\author{Ahmed Katary}
\email{akatary@mit.edu}
\orcid{0009-0003-0034-5316}
\affiliation{%
  \institution{MIT CSAIL}
  \city{Cambridge}
  \state{MA}
  \country{USA}
}

\author{Tarik Hasic}
\email{thasic@mit.edu}
\orcid{0009-0007-0734-1157}
\affiliation{%
  \institution{MIT CSAIL}
  \city{Cambridge}
  \state{MA}
  \country{USA}
}

\author{Amira Abdel-Rahman}
\email{amira.abdel-rahman@cba.mit.edu}
\orcid{0000-0002-7183-8818}
\affiliation{%
  \institution{Center for Bits and Atoms, MIT}
  \city{Cambridge}
  \state{MA}
  \country{USA}
}

\author{Nayeemur Rahman}
\email{nayeem31@mit.edu}
\orcid{0009-0007-1273-4094}
\affiliation{%
  \institution{MIT CSAIL}
  \city{Cambridge}
  \state{MA}
  \country{USA}
}
\author{Leandra Tejedor}
\email{leandra0@mit.edu}
\orcid{0009-0004-5153-830X}
\affiliation{%
  \institution{MIT CSAIL}
  \city{Cambridge}
  \state{MA}
  \country{USA}
}
\author{Mackenzie Leake}
\email{leake@mit.edu}
\orcid{0000-0002-8070-4918}
\affiliation{%
  \institution{MIT CSAIL}
  \city{Cambridge}
  \state{MA}
  \country{USA}
}
\author{Megan Hofmann}
\authornote{Equal contribution}
\orcid{0000-0003-2283-8587}
\affiliation{%
  \institution{Khoury College of Computer Sciences, Northeastern University}
  \city{Boston}
  \state{MA}
  \country{USA}
}
\author{Stefanie Mueller}
\email{stefanie.mueller@mit.edu}
\orcid{0000-0001-7743-7807}
\authornotemark[1]
\affiliation{%
  \institution{MIT CSAIL}
  \city{Cambridge}
  \state{MA}
  \country{USA}
}

\renewcommand{\shortauthors}{Faruqi et al.}



\begin{abstract}
With recent advances in Generative AI, it is becoming easier to automatically manipulate 3D models. However, current methods tend to apply edits to models globally, which risks compromising the intended functionality of the 3D model when fabricated in the physical world. For example, modifying functional segments in 3D models, such as the base of a vase, could break the original functionality of the model, thus causing the vase to fall over. We introduce a method for automatically segmenting 3D models into functional and aesthetic elements. This method allows users to selectively modify aesthetic segments of 3D models, without affecting the functional segments. To develop this method we first create a taxonomy of functionality in 3D models by qualitatively analyzing 1000 models sourced from a popular 3D printing repository, Thingiverse. With this taxonomy, we develop a semi-automatic classification method to decompose 3D models into functional and aesthetic elements. We propose a system called \textit{Style2Fab} that allows users to selectively stylize 3D models without compromising their functionality. We evaluate the effectiveness of our classification method compared to human-annotated data, and demonstrate the utility of Style2Fab with a user study to show that functionality-aware segmentation helps preserve model functionality.

 \end{abstract}

%





\begin{teaserfigure}
\centering
  \includegraphics[width=\textwidth]{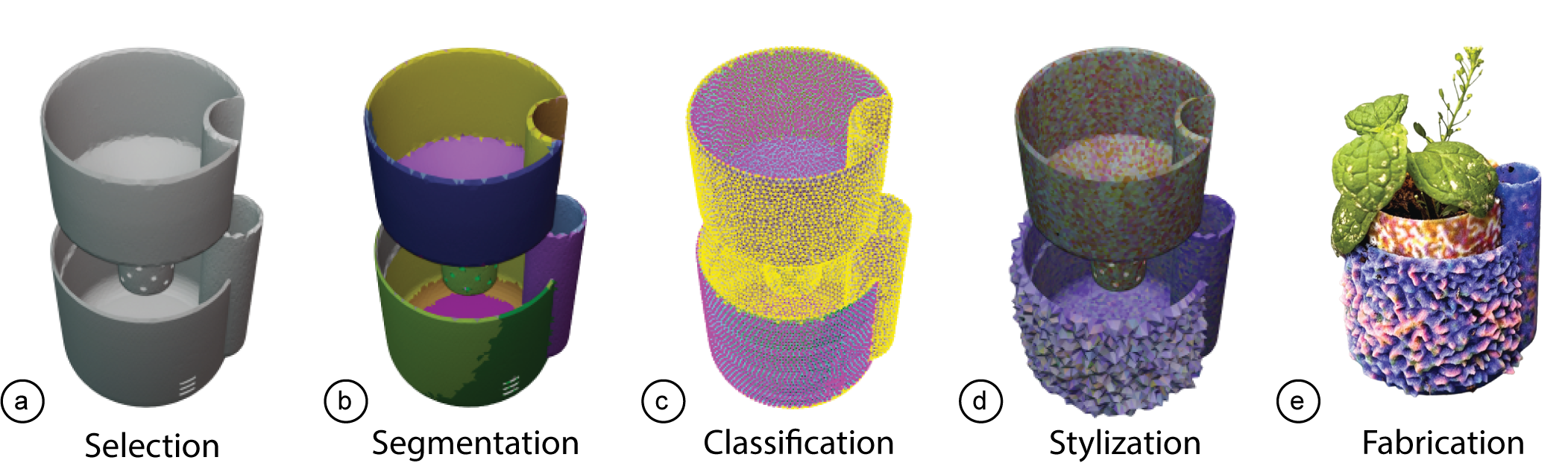}
  \vspace{-10pt}
  \caption{To stylize a 3D model with Generative AI without affecting its functionality a user: (a)~selects a model to stylize, (b)~segments the model, (c)~automatically classifies the aesthetic and functional segments, (d)~selectively styles only the aesthetic segments, and (e)~fabricates their stylized model.}
  \label{fig:teaser}
\end{teaserfigure}

\maketitle

\section{Introduction}

   A key challenge for many makers is modifying or ``\textit{stylizing}''~\cite{oehlberg2015patterns, alcock2016barriers} open source designs shared in online repositories \cite{buehler2015sharing, Kuznetsov_2010_expertamateur} (e.g., Thingiverse \cite{Thingiverse2023Apr}). While these platforms provide numerous ready-to-print 3D models, customization is limited to changing predefined parameters \cite{alcock2016barriers}. While recent advances in deep-learning methods enable aesthetic modifications in 3D models with \textit{styles} \cite{michel2022text2mesh, siddiqui2022texturify,gao2023textdeformer}, customizing existing models with these styles presents new challenges. Beyond aesthetics, 3D printed models often have designed functionality that is directly related to geometry. Manipulating an entire 3D model, which can change the whole geometry, may break this functionality.  Styles can be selectively applied, but this requires the maker to identify which pieces of a 3D model affect the functionality and which are purely aesthetic --- a daunting task for users remixing unfamiliar designs. In some cases,  users can label functionality in CAD tools~\cite{hofmann2018greater, veuskens2020coda}, however, most of the models shared in online repositories are 3D models that have lost this key meta-data.
   
To help makers make use of emerging AI-based 3D manipulation tools, we present a method that automatically decomposes 3D meshes designed for 3D printing into components based on their functional and aesthetic parts. This method allows makers to selectively stylize 3D models while maintaining the desired original functionality. Derived from a formative study of 1000 designs on the Thingiverse repository, we contribute a taxonomy for classifying geometric components of a 3D mesh as (1) \textit{aesthetic}, contributing only to model aesthetics; (2) \textit{internally}-functional, related to assembly of component-based models; or (3) \textit{externally}-functional, related to an interaction with the environment. Based on this taxonomy, we contribute a topology-based method that can automatically segment 3D meshes, and classify the functionality of those segments into these three categories. To demonstrate this method, we present an interactive tool, ``Style2Fab'', that enables makers to manipulate 3D meshes without modifying their functionality. Style2Fab uses differentiable rendering~\cite{kato2020differentiable} for stylization as proposed in Text2Mesh~\cite{michel2022text2mesh}. Our work demonstrates how we extend these methods to enable complex manipulation of open-source 3D meshes for 3D printing without modifying their original functionality.

\changes{Consider a leading scenario where an inexperienced maker, Alex, wants to stylize the outside of a 3D printable self-watering planter \autoref{fig:teaser}. Conceptually, Alex understands that the base needs to stay flat and the interlocking segments of the two components of the model should remain unchanged. But she does not know how to isolate these regions in the two 3D meshes. She processes the models in Style2Fab, where our functionality-aware segmentation method segments the two models, and labels the base and interlocking segments as \textit{functional}. Our method has done the work of tediously editing the model for her, allowing her to verify that the model's  functionality was preserved on the segment level. She applies her style only to the outer edge of the planters and sends it off to the 3D printer. She uses the final design to grow herbs on her desk. }

In the remaining sections, we survey the literature on tools that support makers in modifying 3D printable models and present a formative study on 3D model functionality in the context of 3D printed designs shared online. We then present our method for automatically segmenting and classifying the functional components of 3D models. Next, we present the Style2Fab system which uses this segmentation method to help makers stylize 3D printable designs. We use Style2Fab to evaluate if our classification and segmentation method can help makers modify existing designs without breaking their functionality.

\changes{
\section{Related Work}


To situate our findings and proposed system, we draw upon research on open-source 3D designs, support tools for 3D modeling and printing, systems for functionality-aware design, and data-centric methods for 3D manipulation. 

\subsection{Personalizing Open-Source 3D Designs}
As 3D printing emerged as expert-amateur makers' digital fabrication tool of choice~\cite{Kuznetsov_2010_expertamateur}, they began to share their 3D models and designs online in open repositories. Alcock~\etal~\cite{alcock2016barriers} argue that this makes online repositories like Thingiverse ideal training grounds for novice makers. Numerous studies have explored the practices of these novice makers~\cite{hudson2016understanding, norouzi2021making, berman2021howdiy, schmidt2013design} and tend to find that makers struggle to make any modifications to existing designs because of the limitations of current computer-aided design (CAD) workflows.  Unfortunately, the types of designs shared (e.g., 3D meshes) are largely incompatible with the capabilities of these novices. Tools for editing 3D printable meshes require advanced expertise~\cite{schmidt2010meshmixer}. Oehlberg \etal~\cite{oehlberg2015patterns} observed that, even when designs are customizable, they lack the full range of modifiable features makers seek. 

\subsection{Supporting users in Fabrication Workflows}

One approach to address this issue is to build on the model formats makers are more likely to share: 3D meshes. Researchers have developed a variety of unique tools with that approach. Grafter~\cite{roumen2018grafter} automates the process of recombining mechanical elements of 3D printed mechanisms. Similarly, AutoConnect~\cite{koyama2015autoconnect} presents an automatic method for creating 3D printable connectors to attach two physical objects together. RetroFab~\cite{ramakers2016retrofab}, Makers' Marks~\cite{savage2015makers}, and Robiot~\cite{li2019robiot} offer different approaches to modifying physical interfaces. Reprise~\cite{xiang_anthony_chen_reprise:_2016} takes a domain-focused approach and analyzes the features of assistive 3D printed devices to create remixable patterns. Alternatively, systems like Encore~\cite{chen2015encore} and Medley~\cite{chen_medley} offer ways to integrate real-world objects into existing 3D models. 

Several optimization algorithms target post-fabrication physical properties of 3D models, such as balance~\cite{prevost2013make}, moment-of-inertia for spinnable models~\cite{bacher2014spin}, or constrained optimization between center-of-mass and balance~\cite{zhao2016make, prevost2016balancing}. Other proposed systems provide users with design capabilities and existing functionality, such as adding helical springs and joints to allow deformation and extension~\cite{he2019ondule}, designing hand-launched free-flight glider airplanes~\cite{umetani2014pteromys}, and embedding rotary, linear and chained mechanisms within 3D objects for laser cutting~\cite{leen2020lamifold}. 

\subsection{Functionality-Aware Design in 3D Models}

Another approach to addressing the barrier to personalizing open-source designs is to prompt makers to produce better models with meta-data specific to fabrication. Making modifications on 3D models with the intent of fabricating them can be a complex process since the modifications have to satisfy fabrication constraints. Schmidt \etal~\cite{schmidt2013design} proposed an approach called `Design to Fabricate' with an aim towards such modifications. The related tool Meshmixer~\cite{schmidt2010meshmixer} allows users to manipulate 3D models with the intent to 3D print them afterwards. Shugrina \etal~\cite{shugrina2015fab} define a `Fab Form' as any design representation that lends itself to interactive customization by a novice user, while remaining valid and manufacturable. In their proposed system, they achieve these requirements for general parametric designs (previously tagged) with a general set of automated validity tests and exposing a small number of parameters to the user. Schulz \etal~\cite{schulz2014design} enable makers to combine parameterized template parts from a database to create new models.  CODA~\cite{veuskens2020coda} extends this work by proposing an interactive support tool that communicates the implications of parametric manipulations on CAD models to users, based on static and dynamic constraints from the model. Other systems present interactive tools and algorithms to design functionality-aware objects. Zhang et al.~\cite{zhang2017functionality} propose an approach to retarget existing mechanical templates to user-specific input shapes using parameterized mechanisms. Hofmann et al.~\cite{hofmann2018greater} present a system that allows modelers to graphically specify their design intents by associating 3D model geometry with small snippets of code that can modify or evaluate that geometry as it is edited in the 3D modeling environment. 

However, the majority of the models shared online~\cite{alcock2016barriers} are in the form of 3D meshes, which do not always have the necessary meta-data required to allow such parametric manipulations. Given this constraint, an approach to allow personalization of shared models is to segment models into elements related to `aesthetics' and those related to `function'. Post-segmentation, users can manipulate the `aesthetic' components while preserving the functional segments. Laga \etal~\cite{laga2013geometry} present a heuristics-based method to find semantic correspondences between segments in 3D models for four classes of models (Candles, Vases, Chairs, and Lamps), while Lun et al.~\cite{lun2016functionality} present a style transfer method using local substitution, removal, and other operations. Zheng et al.~\cite{mitra_smartVariations13} present an approach for reusing model parts to create new designs by leveraging the symmetric functional substructure between models. Style2Fab extends this line of work by presenting a functionality-aware segmentation approach that leverages segment-level similarity in 3D models to classify functionality. Using this approach, we present a semi-automatic method to separate aesthetic and functional elements in 3D models, allowing users to stylize 3D models without losing functionality.  


\subsection{Data-Centric Methods for 3D Manipulation}


Novel methods such as deep learning leverage larger datasets of 3D models to gain generalizable insights. Hanocka \etal~\cite{hanocka2019meshcnn} proposed \textit{MeshCNN}, a convolutional network to reduce complexity in 3D models based on the observation that 3D models contain redundant information and can be compressed without losing their most important traits; this representation can then be used for tasks such as classification and segmentation.  Yu \etal~\cite{yu2019partnet} used a large, well-annotated PartNet~\cite{mo2019partnet} dataset to identify semantic relationships between different 3D models.

These deep learning based methods can also be used to manipulate the geometry in 3D models. Text2Mesh~\cite{michel2022text2mesh} and TextDeformer~\cite{gao2023textdeformer} leverage CLIP~\cite{radford2021learning} representation to manipulate 3D models based on text prompts, while 3DHighlighter~\cite{decatur20233d} uses CLIP to localize semantic regions on 3D models. Siddiqui \etal~\cite{siddiqui2022texturify}, Suzuki \etal~\cite{suzuki2017autocomplete}, and Yin \etal~\cite{yin20213dstylenet} support applying textures to target surfaces of 3D objects based on multimodal inputs.  Finally, approaches such as Shape-E~\cite{jun2023shap_e} and Magic3D~\cite{lin2023magic3d} allow generation of 3D models with text-prompts. 


The development of these novel approaches has the potential to enhance the current workflow for makers and new creators. However, its important to explore which component of this process should be augmented with these novel techniques. In this paper, we focus on stylizing 3D models and explore how data-centric methods can help users separate functional from aesthetic elements, allowing them to stylize the aesthetic parts of the 3D models without compromising functionality.    

}

\section{Formative Study}
\begin{figure*}[!ht]
    \includegraphics[width=\textwidth]{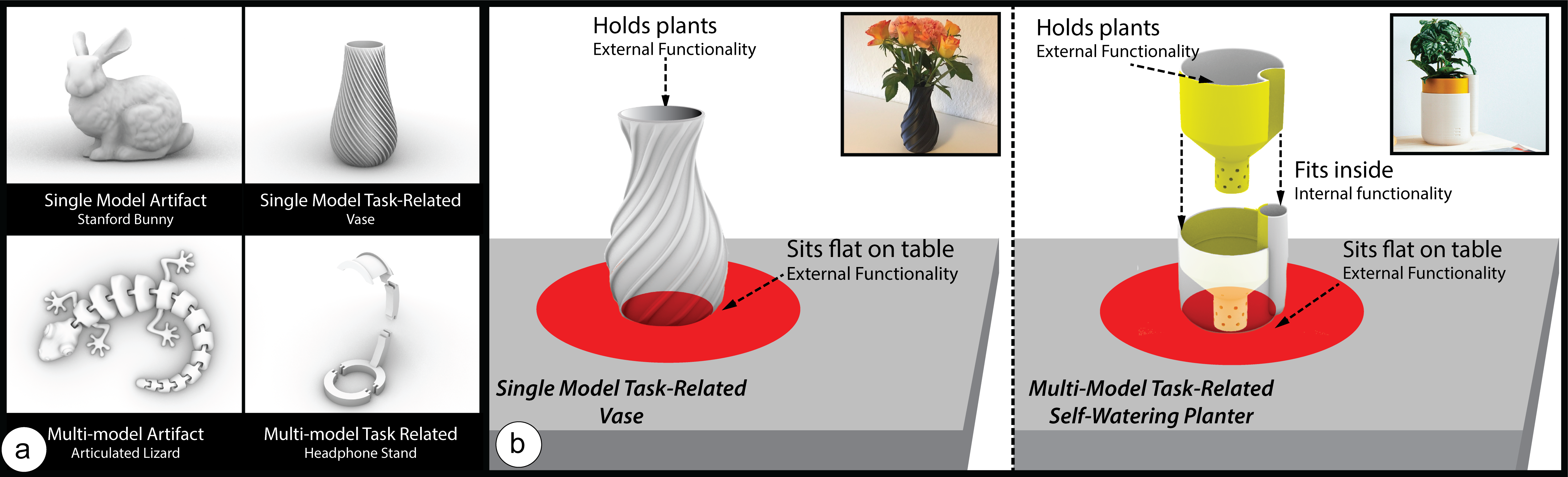}
    \caption{Categories of 3D models based on functionality: (a)~We identify four categories of models defined by two dimensions: Artifact vs Task-Related and Single vs Multi-Component models. These dimensions align with differences between external and internal contexts, (b)~shows an example of external and internal functionality on segments of a vase and a self-watering planter.  }
    \label{fig:formative_study_models}
\end{figure*}

One of the key challenges in modifying 3D printable models is ensuring that they remain functional. This requires a maker to carefully identify which parts of a design contribute to the functionality and which parts contribute only to the model's aesthetics. The aim of this formative study is to identify functionality descriptors in a wide variety of 3D models. To do this we qualitatively coded 1000 designs sourced from Thingiverse using a similar approach to Hofmann~\cite{hofmann2018greater} \etal and Chen \etal~\cite{xiang_anthony_chen_reprise:_2016}. From these codes, we developed a taxonomy of 3D model functionality.

\subsection{Data Collection}
Thingiverse is a popular online resource for novice and expert makers to share 3D printing designs or \textit{things}\footnote{A \textit{thing} refers to a design that may include many 3D meshes. We use the term model to refer to one 3D mesh file rather than all of the independent components shared in a \textit{thing}.}. While some models are shared in editable formats, most are shared as difficult-to-modify 3D meshes in OBJ and STL file formats (OBJ/STL). We collected and analyzed the 1000 most popular\footnote{The popularity metric is determined by Thingiverse which aggregates downloads, shares, likes, and remix counts.} \textit{things} on Thingiverse as of January 23rd, 2023. \changes{ Although large-scale datasets exist for segmentation tasks (COSEG~\cite{wang2012active} and PartNet~\cite{mo2019partnet}), we found that sampling models from Thingiverse provides a wider variety of 3D models suited for fabrication since existing data sets are not intended for 3D printing.}

We organized and standardized all 3D models included in these 1000 \textit{things}. We first excluded any 3D models that were not in an STL, OBJ, or SCAD format, limiting our data to 3D meshes. We excluded all corrupted 3D models, and the ones shared without the three given formats. We converted all remaining models to the OBJ format. Next, we manually excluded duplicate meshes of varied sizes since this would not contribute to our classification of functional components. For highly-similar models shared as a collection (e.g., `Fantasy Mini Collection', Thing ID: 3054701), we kept only a single mesh. After this preprocessing, we had a total of 993 different Thingiverse \textit{things}, comprising 10,945 unique 3D models (i.e., objects or parts of objects).  

\subsection{Inductive Taxonomy Development}
We used an iterative qualitative coding method to develop our taxonomy of functionality. We first inductively coded 100 randomly selected models to develop an understanding of the functionality of these 3D-printed models. After negotiation across coders, we identified two distinct categories of 3D models based on their functionality: \textit{Artifacts} and \textit{Task-Related Models}. Artifacts are objects that serve primarily aesthetic purposes, such as statues. Task-Related models have been designed to help perform a specific task, such as a phone stand or battery dispenser. Both Artifacts and Task-Related models can be composed of single or multiple components assembled together~\autoref{fig:formative_study_models}.

From this classification of types of models, we determined two axes that we can use to classify if a segment of a 3D mesh can be modified without changing the intended functionality of the design. The first axis is \textit{external context}, which describes how the surface of the segment interfaces with the real world to affect the functionality of the model (e.g., the flat base of a planter interfaces with a table surface). Most Artifacts have few segments with external contexts, while Task-Related models have many. The second axis is internal context; a segment has high internal context if it interfaces with other segments within the same thing to affect the design's functionality (e.g., linkages in an articulated lizard). Segments that do not have internal or external context are considered \textit{aesthetic} since they do not affect functionality. Segments without internal and external context can readily be modified since they only serve an aesthetic purpose.

\subsection{Deductive Functionality Classification} 
Using this taxonomy, we then labeled our entire data set of 993 models based on the two types of designs and the axes of internal and external context. For each 3D model, two annotators examined the associated Thingiverse meta-data to understand the intended functionality of the model using shared images of the model being used in specific scenarios. Independently, each annotator labeled the model as \textit{Artifact} or \textit{Task-Related} based on its external functionality. This resulted in a Cohen's Kappa inter-rater reliability score of 0.94. \changes{They negotiated differences to finalize the labels for each model, reaching full agreement. Two examples of models that required negotiation are ThingID:3096598 (Chainmail) and ThingID:1015238 (Robotic Arm). The annotators resolved the former to be an Artifact because the Thingiverse page did not showcase any task-specific use case, and the latter to be Task-Related since its metadata contained videos describing a specific functionality.}
Following data classification, we removed all models that had no aesthetic segments because these cannot be readily modified. This exclusion primarily removed models used for calibrating printers where any change would have changed the functionality. Our final, labeled data set contained 938 models and is summarized in~\autoref{tab:model_class}.

\begin{table}[ht!]
\centering

\resizebox{\linewidth}{!}{%
\begin{tabular}{@{}crrr@{}}
                 & \multicolumn{1}{c}{\textbf{Single-Component}} & \multicolumn{1}{c}{\textbf{Multi-Component}} & \multicolumn{1}{c}{\textbf{Total}} \\ \hline
\textbf{Artifact} & 46                                     & 320                                       & \textbf{366}                       \\
\rowcolor{lightgray}
\textbf{Task-Related}       & 91                                     & 481                                       & \textbf{572}                       \\ \midrule
\textbf{Total}            & \textbf{137}                           & \textbf{801}                              & \textbf{938}   \\                   
\end{tabular}%
}
\caption{Counts of Thingiverse designs based on dimensions of internal and external functionality. Rows reflect external context and columns reflect internal context.}
\label{tab:model_class}
\vspace{-10pt}
\end{table}



\section{Functionality-Aware Segmentation}
Our formative study helps us better understand the types of functional segments that affect the functionality of the model. Based on these results and our data set, we present a method for functionality-aware segmentation and classification of 3D meshes designed for 3D printing. In this section, we first define our segmentation problem and present our segmentation approach. Next, we present a method for classifying internal and external functionality on each segment of a model. We present our approach to tuning important hyper-parameters that affect the efficacy of our method and an evaluation of our classification approach compared to labels generated in our formative study. Finally, we present our Style2Fab user interface developed using this functionality-aware segmentation approach.


We use an unsupervised segmentation method based on spectral segmentation that leverages the mesh geometry to predict a mesh-specific number of segments. This method allows us to generalize across 3D printable models with diverse functionality. Using our Thingiverse data set, we evaluated this method for its accuracy in predicting the number of functional segments and its ability to handle a wide range of mesh resolutions.

\subsection{Segmentation Approach}
The process of segmenting a 3D mesh can be defined as finding a partition of the mesh such that the points in  different clusters are dissimilar from each other, while points within the same cluster are similar to each other.
We use a spectral segmentation process that leverages the spectral properties of a graph representation of the 3D mesh to identify meaningful segments. By examining the eigenvectors and eigenvalues of the graph norm Laplacian matrix, this method captures the underlying structure of the mesh and groups similar vertices together, resulting in a meaningful partition of the model.

Consider a 3D mesh as a graph where nodes represent a set of faces and edges represent connections between adjacent faces. The segmentation problem is to decompose the mesh into $k$ non-overlapping sub-graphs that represent a piece of the model with consistent features (e.g., the base, outer rim, and inside of a vase). The hyper-parameter $k$ can be any integer value between 1, one segment containing all faces in the mesh, and $n$, one segment for every individual face in the mesh. If $k$ is too low, the segments will not be able to isolate components with unique functionality (e.g., the base of a vase is not separate from the outside). \changes{ If $k$ is too high, functional components of the model may be split into multiple segments and may be modified in incompatible ways (e.g., half of the base is stylized with a surface texture and the other half has the original flat surface). A key challenge in functionality-aware segmentation is automatically selecting a value of $k$ for each design; we do not assume that makers will be able to easily identify a good value of $k$ when examining a design.}

\subsubsection{Predicting the Number of Segments}

We use a heuristic-based approach for estimating a value of $k$ that partitions a mesh into segments that isolate functionality. Using a 3D mesh's degree- and adjacency-matrix, we use spectral decomposition~\cite{katz2003hierarchical} to extract an eigenbasis for the mesh. This allows us to use the resulting eigenvalue distribution, representing the connectedness of a mesh, to identify a partition yielding the highest connectedness for individual segments.

We first describe the spectral segmentation approach. Given a 3D mesh where $F$ represents the set of faces, we first construct a weighted adjacency matrix $W$. The element $W_{ij}$ represents the similarity between faces $f_i$ and $f_j$, calculated using the shortest geodesic distance between the centers of faces $f_i$ and $f_j$ and the angular distance between them as defined by~\cite{katz2003hierarchical}. 

We use the weight matrix, $W$, and the degree matrix, $D$ in order to compute the eigenvectors and values of the face graph. Formally defined as the norm Laplacian of a graph, $L = \sqrt{D}^{T}W\sqrt{D}$. From the eigenvalues of $L$, we are able to capture the connectedness of the mesh, where large gaps between eigenvalues imply weak connectedness. 

    

In the approach proposed by Liu and Zhang~\cite{liu2004segmentation}, the eigenvectors corresponding to the smallest $k$ eigenvalues $\lambda$ are used to construct a k-dimensional feature space, where $k$ is the desired number of segments. Instead of using the smallest $k$ eigenvalues, we analyze the entire distribution of eigenvalues $\lambda$. A high standard deviation in the eigenvalue distribution indicates that the eigenvalues are spread out over a wide range, which suggests a more complex graph structure with varying connectivity and potentially multiple distinct clusters or segments. In this case, the graph may benefit from a more refined segmentation process. On the other hand, a low standard deviation implies that the eigenvalues are more tightly clustered, which suggests a relatively uniform graph structure with fewer distinct clusters. In this case, it would be sufficient to partition the graph into a lower number of clusters. We leverage this distribution to automatically calculate a value of $k$. Specifically, we calculate the number of eigenvalues that have a higher dispersion than the distribution's standard deviation, using~\autoref{eigengap-calculation}.
\begin{equation}
    k = \left|\left\{ \lambda_i : \lambda_i > \mu + \sigma, i = 1, \dots, n \right\}\right|
    \label{eigengap-calculation}
\end{equation}


Once we have extracted the lowest $k$ eigenvectors and their corresponding eigenvalues we follow Liu and Zhang's~\cite{liu2004segmentation} segmentation method that uses k-means clustering to identify segments spanning from the $n$ faces captured by these high-variation eigenvectors. Based on the resulting clusters, we assign each face in the mesh graph to its corresponding segment, resulting in a segmented 3D model.

\subsubsection{Uniform Mesh Resolution for Segmentation} 
This segmentation approach is dependent on a uniform resolution of a mesh; non-uniform meshes will produce incoherent segmentation as some portions of the model are represented by too few faces and other portions have too many faces. Unlike other segmentation approaches, our data set of real-world models did not have consistent resolution and this would have affected the utility of our method. Thus, we re-mesh all models to give them a uniform 25k resolution using Pymeshlab~\cite{pymeshlab}. Note that this process' runtime increases with the resolution. Therefore, we want a low-resolution value that does not negatively affect our segmentation and classification method. 

\begin{figure}[ht!]
    \includegraphics[width=\linewidth]{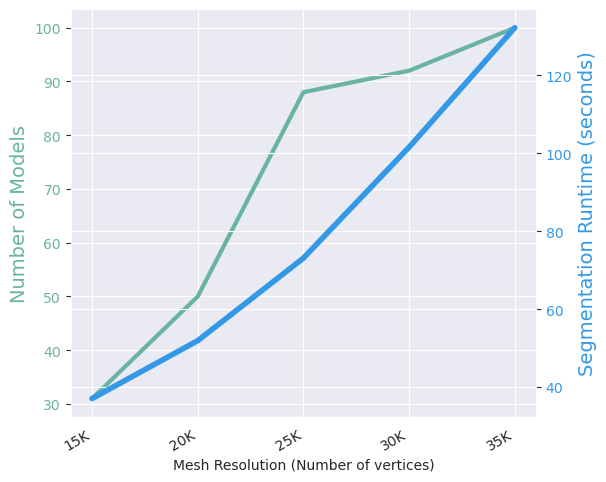}
    \caption{Comparison plots of the number of stabilized models (green) and segmentation time (blue) against mesh resolution.}
    \label{fig:model_stability}
\end{figure}

We determined that resolution to be 25k (vertices) by segmenting and comparing 100 randomly selected 3D models from our data set. For each mesh, we segmented the remeshed models with 15K, 20K, 25K, 30K, and 35K vertices.  We then looked for the lowest resolution that stabilized the predicted number of segments $k$. That is, for all higher resolutions, the number $k$ did not change. For 88\% of models, a 25k resolution stabilized this value. The segmentation at 25K resolution took an average of 72 seconds, while segmentation at 30K resolution took an average of 102 seconds. \autoref{fig:model_stability} shows the effect of mesh resolution on the number of models it stabilized and the time it took to complete segmentation. 


\subsection{Analyzing Functionality in 3D Models}
\label{section:functionality-classification-method}
After segmentation, the system must classify each segment as \textit{functional} or \textit{aesthetic}. To do this we use a heuristic that infers that if a segment $i$ is topologically similar to (i.e., shaped like) another segment $j$, the functionality of $i$ will be the same as $j$. Thus, to classify each segment, we must find some similar topology that has already been labeled as functional. Using the taxonomy of internal and external functionality from our formative study, we can break up the problem of finding a similar, labeled segment $j$ into two approaches: (1) we analyze external functionality by identifying topologically similar segments in models in our Thingiverse data set, and (2) we determine internal-functionality in multi-component models by identifying linkages between components based on topologically similar segments. However, it is critical to compare segments and their parent meshes because comparing only segments introduces noise. Segments without the context of their parent mesh are just geometrical features that, while topologically similar, are likely to be used in different ways. 

\subsubsection{Using Similarity as a Heuristic:} 
\label{similarity_subsection}
We hypothesize that similar models will use similar geometries to enact a similar functionality.  We use the approach of measuring topological similarity from Hilaga \etal~\cite{hilaga2001topology}, which uses a Multiresolution Reeb Graph representation (MRGs) of meshes to analyze the similarity. This method is ideal for our domain due to its invariance to translation, robustness to connectivity variations, and computational efficiency. Note that we can use this method on both whole meshes and individual segments since a segment is, itself, a mesh of connected faces.  Thus, given segment $s_i$ in a mesh $m_i$ and segment $s_j$ in mesh $m_j$, the contextual similarity $Contextual\_Sim(s_i, s_j)$ is the product of the similarity between the segments and their parent meshes~(\autoref{eq:sim}). This gives us a value between 0 (i.e., a complete topological mismatch) and 1 (i.e., identical topology).

\begin{equation}\label{eq:sim}
    Contextual\_Sim(s_i, s_j) = sim(m_i, m_j) \cdot sim(s_i, s_j)
\end{equation}

Given a segment $s$ in mesh $m$ and a set of other meshes $\mathbb{M}$, we can find the similarity between $s$ and all of the segments $\mathbb{S}$ in the other meshes. From these similarity values, we decide the label of $s$'s functionality by comparing it to the labels on the subset of $\mathbb{S}$ that are most similar $\mathbb{S}_{sim}$. We take a uniformly weighted vote of the labels on each segment in $\mathbb{S}_{sim}$ and classify $s$ as the majority label. Regardless of the similarity of these most similar segments, we weigh their label votes equally. We empirically found that the accuracy of functionality classification converges after comparing $s$ to five other similar segments (i.e., $|\mathbb{S}|=5$). 

Now that we have a method of labeling a segment based on a related set of pre-labeled meshes $\mathbb{M}$, we must find mesh sets that help us to identify, separately, internal and external functionality. The size of $\mathbb{M}$ will have a significant effect on the time it takes to compute segment similarity. For each mesh $m_i$ in the set, there will be $k_i$ segments to compare to $s$. Thus, as the size of the mesh set increases the number of similarity comparisons increases by a factor of $k_i$ and quickly becomes too time-consuming to compute. To classify functionality, we need to find a small set $\mathbb{M}$ that provides the most information about the segment with the least amount of noise. Our insights into the differences between internal and external functionality help us to select good sets of meshes.


\subsubsection{Classifying External Functionality}
To identify external functionality, we compare segments to models that function similarly using similar geometric features. We built a labeled data set of segmented models from Thingiverse for identifying external functionality from the 46 \textit{Artifacts} and 91 \textit{Task-Related} single-component models in our data set. First, we segmented all of these models using our segmentation method and produced 1151 different segments. Then two annotators analyzed each segment with contextual information from the parent model's Thingiverse page and independently labeled the segment as \textit{Aesthetic} or \textit{Functional}. We also asked the annotators to independently label a segment if it contained an \textit{Aesthetic} and \textit{Functional} component fused together (inefficient segmentation). After reviewing all segments, they had an inter-rater reliability of 0.97. They negotiated all disagreements to produce our ground truth classification of segment type. At the end of the study, 51\% of the segments were annotated as \textit{Aesthetic}, while \textit{49\%} segments were annotated as \textit{Functional}. From the functional segments, the annotators agreed that 24 segments (2\%) from 17  different (12.4\%) models were composed of \textit{Aesthetic} and \textit{Functional} elements fused together, leading them to annotate the entire segment as \textit{Functional}.

Naively, we could classify the external functionality of a segment $s$ in a mesh $m$ by comparing it to all meshes in this data set. However, this would be computationally expensive and introduce noise since segment-level similarity may occur between segments that are used in different ways in models with different uses. Thus, we prune the set of labeled models by first comparing mesh-to-mesh similarity. Thus, we collect the set of most-similar meshes $\mathbb{M}_{sim}$ from our data set. As shown in Section~\ref{similarity_subsection}, we found that five meshes were sufficient. Considering External Functionality to be a binary label, we got a Precision score of 74\% and a recall of 88\%. Hence, this is a conservative approach towards functionality prediction, where the classifier has a higher false positive rate. It is more important that the system does not modify functional components than that it misses aesthetic components. 

\subsubsection{Classifying Internal-Functionality}
Similar to the case of external context, we can identify internal functionality by comparing a segment to a set of segmented models with similar internal functionality. However, internal context is dependent on similarity within a group of meshes that make up a mechanism. Two segments have internal functionality if they interface with each other to form a linkage in a mechanism. Thus, our problem was to identify pairs of segments between components (multiple models that compose a single mechanism) and classify them as internally functional. We approach the problem in the following manner: We first create the set $(Combinations(\mathbb{M}_m)$) containing all possible pairs of segments not from the same mesh. For any model $m$, we can now search over all possible pairs of segments between different meshes from the set $Combinations(\mathbb{M}_m)$, and identify similar segments.

In this case, instead of assigning the label by taking a vote across the five most similar segments (i.e., $\mathbb{S}_{sim}$), we decide that there is a linkage between a segment $s$ and another segment $s_c$, in another component mesh, if they have a similarity value greater than the hyper-parameter $\alpha$. In the case that multiple segments have high enough similarity, $s_c$ will be the segment with the highest similarity to $s$. Once identified, we label both $s$ and $s_c$ as having internal functionality.

To increase efficiency, we prune the number of segment comparisons in this process. First, we can exclude all segments in meshes that have been labeled as having external functionality. Ultimately, we are looking for a single binary label between aesthetic and functional, so it does not matter if a segment is both internally and externally functional. Second, as we identify linkages we can remove both $s$ and $s_c$ from future comparisons within the mechanism because linkages are formed of only two segments. 

To decide that two segments are sufficiently similar to be considered a linkage, we need a threshold similarity value of $\alpha$. To identify this threshold we gathered ground truth data from our data set. Two researchers annotated segments of the multi-component models in our data set to identify linked segments. We randomly selected 50~things with multiple components which contained a total of 157~component meshes. For each model, two annotators independently labeled all pairs of segments that formed a linkage in the mechanism resulting in an inter-rater reliability score of .99. They negotiated disagreements and produced a ground truth dataset. From this data set, we identified an effective threshold similarity score $\alpha~=0.86$ by evaluating the precision and recall across multiple $\alpha$ values. We found that $\alpha~=0.86$ maximized the identification of functional segments and then minimized the misidentification of aesthetic segments. By this analysis, we got a precision value of 64\% and a recall value of 86\%. Like classifying external context, we prioritize high recall over precision since the cost of missing a functional segment will break a model while missing an aesthetic segment will only affect aesthetics. Therefore, we opted to have a more conservative classifier for both the internal and external functionality. 



\subsection{Stylization of Segments}

\begin{figure}[t]
    \centering
    \includegraphics[width=0.7\linewidth]{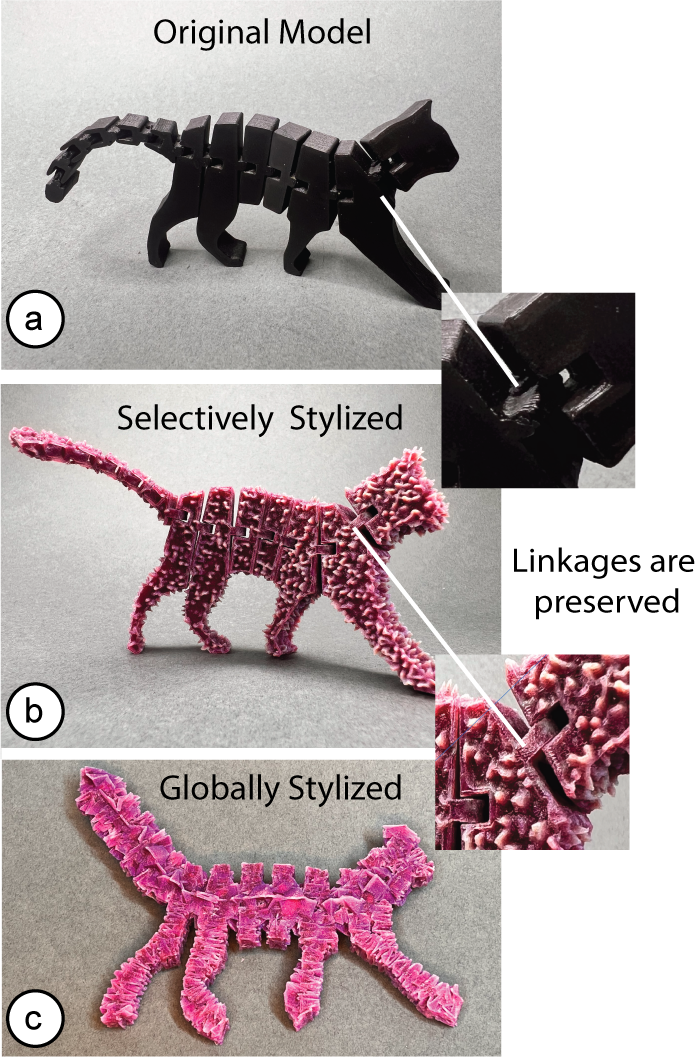}
    \caption{A demonstration of the differences between functionality-aware and global styling. A ``flexi cat'' model (Thing: 3576952) is shown (a) without styles, (b) with functionality-aware styles, and (c) functionally broken by global styles.}
    \label{fig:global_selective}
\end{figure}

We use Text2Mesh~\cite{michel2022text2mesh} to stylize models based on text prompts. Text2Mesh uses a neural network architecture that leverages the CLIP~\cite{radford2021learning} representation. The system considers a 3D model as a collection of vertices, where each vertex has a color channel (RGB) and a 3D position that can move along its vertex normal. Text2Mesh reduces the loss between the 3D model rendering (CLIP representation) from different angles and the CLIP representation of the textual prompt using gradient descent. Text2Mesh makes small manipulations in both the color channel and vertex displacement along the vertex normal for each of the vertices in order to make it look more similar to the text prompt. This allows the system to generate a stylized 3D model that reflects the user's desired style.

This method will stylize the whole mesh and change the topology of the functional segments. In~\autoref{fig:global_selective}a-c, we show that global stylization can render a functional object, in this case, an articulated cat, inoperable. We augment this system by adding an additional step of masking the gradients and setting functional vertices to zero. This allows manipulation of the color and displacement channels while preserving the functional segments of the model. As specified in Text2Mesh~\cite{michel2022text2mesh}, we run this optimization for 1500 iterations.  

\subsection{Style2Fab User Interface and Workflow}
\begin{figure}[ht!]
    \includegraphics[width=\linewidth]{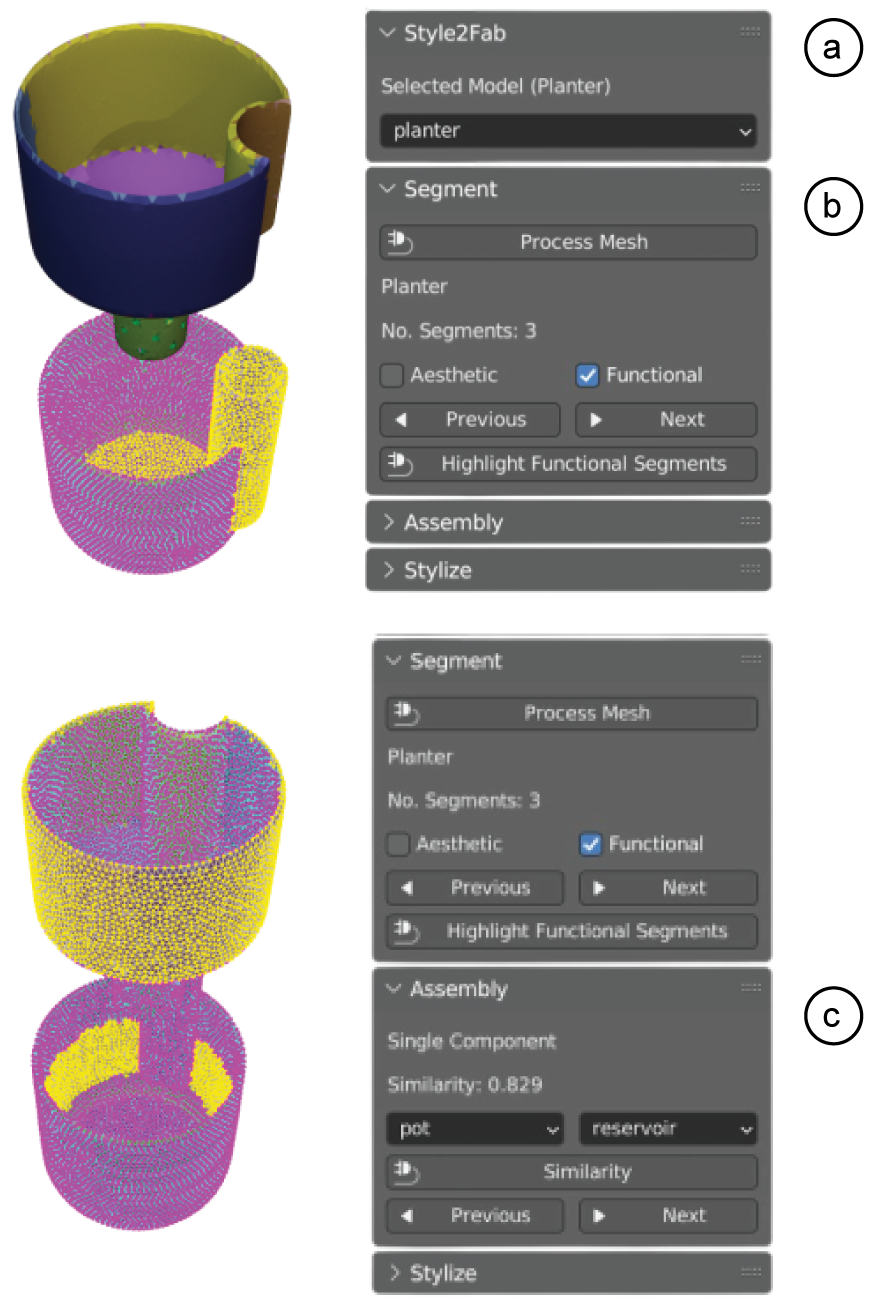}
    \caption{In the Style2Fab UI, the user (a) loads their model and (b) browses the functionality labels after segmentation. (c) In the case of multi-component models, the user can examine and adjust linked segments between components.}
    \label{fig:style2fab_interface-1}
\end{figure}

Style2Fab is a plugin for the open-source 3D design software tool Blender. To stylize a model with Style2Fab, the user must: (1) pre-process their model for segmentation and stylization, (2) segment and classify the functionality of each segment, (3) selectively apply a style to segments based on functionality, and  (4) review their stylized model. We break these tasks up into four menus in the user interface~(\autoref{fig:style2fab_interface-1}). 

\subsubsection{Pre-Processing and Segmentation}
Once the user has loaded an OBJ file of their 3D mesh into the plugin, they ``Process'' the model to standardize its resolution and automatically detect the number of segments (i.e., $k$) needed to classify functionality across the model~(\autoref{fig:style2fab_interface-1}b). By default, the resolution is set to 25k faces based on our evaluation. Next, the system will segment the model and give each segment a unique color to help the user visually identify the segments. If the user wants more or fewer segments they can modify the value of $k$ in the interface and re-segment the model. For multi-component models, the user can load multiple meshes representing each component and segment them in parallel. 

\subsubsection{Functionality-Verification of Segments}
After segmentation, the plugin opens panels displaying the type classification of each segment. Users can then review each segment and determine if they agree with the classification. The user's goal is to identify the set of segments that should not be stylized to preserve the desired functionality of the design. To simplify this process, the user can select ``Highlight all functional segments''~(\autoref{fig:style2fab_interface-1}c) to have all segments classified as functional highlighted in the user interface. If they agree with this segmentation, they can move on to stylization. Otherwise, they can individually review all segments. Next, we describe the process for the user to review individual segments. 

\changes{
First, in order to verify externally functional segments, the user can walk through the model's segments  and toggle the functionality class based on their interpretation of the model. When walking over a segmented model via the interface, the segments are highlighted on the model~(\autoref{fig:style2fab_interface-1}b).

When working with multiple components, the user can review the segments that were classified as having internal context based on linkages between components of the model. The user can use the "Assembly" panel~(\autoref{fig:style2fab_interface-1}c) to see pairs of connected segments on distinct models, and click `Separate' for incorrect assignments. If the user disagrees with this classification, they can adjust this similarity parameter $\alpha$ between 0 and 1.
}

\subsubsection{Selective Stylization of Aesthetic Elements}
Post verification of functionality, users can stylize aesthetic segments of a 3D model by entering a natural language description of their desired style and clicking "Stylize Mesh". The completed model is then rendered alongside the original for review. Users can iterate on this process and apply new styles using new text prompts, or re-segment the model as needed.

\section{User Study}
\begin{table}[ht!]
\centering

\begin{tabular}{@{}lllll@{}}
\toprule
\multicolumn{1}{c}{\textbf{Participant}} & \multicolumn{1}{c}{\textbf{Age}} & \multicolumn{1}{c}{\textbf{Gender}} & \multicolumn{1}{c}{\textbf{3D Modeling}} & \multicolumn{1}{c}{\textbf{3D Printing}} \\ \midrule
P1  &  27 & Male & 6 & 4\\
\rowcolor{lightgray}
P2  &  31 & Male & 3 & 1\\
P3  &  27 & Male & 0 & 1\\
\rowcolor{lightgray}
P4  &  24 & Male & 7 & 5\\
P5  &  32 & Male & 4 & 4\\
\rowcolor{lightgray}
P6  &  26 & Female & 7 & 7\\
P7  &  27 & Male & 1 & 0\\
\rowcolor{lightgray}
P8  &  23 & Male & 13 & 15\\ 
\bottomrule
\end{tabular}%
 \caption{Participant demographics and years of experience with 3D modeling and printing. }\label{tab:participants}
\vspace{-10pt}
\end{table}

To evaluate if our functionality-aware segmentation method \textit{supports users in separating functional elements in 3D models} that they did not design, we had  eight university students~(\autoref{tab:participants}) with varied 3D~modeling and printing experience segment and stylize 3D~models from our Thingiverse data set with and without automatic support from Style2Fab.

We randomly selected eight 3D models representing each of our four categories of models based on internal and external context: two single-component Artifacts, two single-component Task-Related models, two multi-component Artifacts, and two multi-component Task-Related models. We segmented each model using our segmentation method and automatically classified the functionality of those segments. Each participant was presented with two models with segmentation and functionality classification (i.e., experimental group) and two models that were only segmented and that they needed to manually classify (i.e., control group). In each condition, the participant received an Artifact and Task-Related model. One of these was always a single component and the other was a multi-component model. We controlled for model-specific and learning effects by giving each participant a different combination of models and conditions. 

We asked participants to classify each segment in each model as functional or aesthetic. In the experimental condition, participants could accept or modify our functionality-aware classification. In the control condition, they had to make a manual classification. After classifying the segments, the users were asked open-ended questions about their experience. They were compensated with \$20 for the hour-long study.

\subsection{Findings}
\begin{figure}[!ht]
    \includegraphics[width=\linewidth]{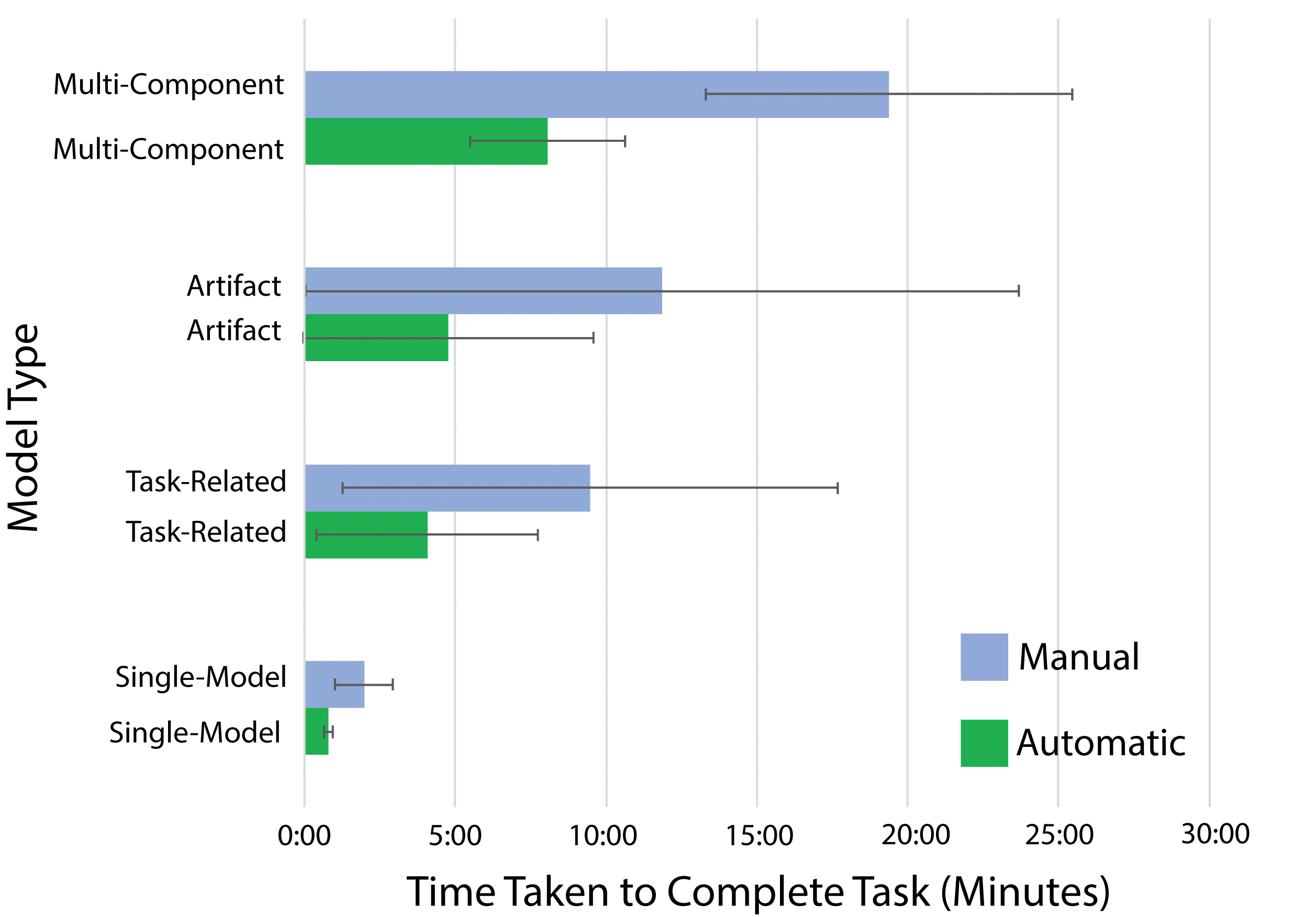}
    \vspace{-10pt}
    \caption{Box plots show the distribution of task completion times by condition. \changes{The more complex the model, the more time users save using the automatic classification of functional segments.}}
    \label{fig:user-study-time}
\end{figure}

In order to understand the differences between the automatic and manual conditions, we compared the time taken to process a model, \changes{which included the classification runtime for the automatic condition,} and the accuracy and precision of the classification~(\autoref{fig:user-study-time}). The time taken by the users was dependent on the complexity of the model, with the single-component model taking the least time, and the multi-component models taking significant time. We conducted a paired one-tailed t-test with 7 degrees of freedom to see if functionality-aware segmentation significantly improved task completion time. Across all types of models, we found a significant improvement in task completion time at the $p<0.05$ level~(\autoref{tab:ttest_time}, \autoref{fig:user-study-time}). The greatest effects were on models with greater complexity such as multi-component models and Task-Related models with multiple segments that had external context.

\begin{table}[ht!]
\centering

\resizebox{\linewidth}{!}{%
\begin{tabular}{@{}cllll@{}}
\toprule
  & \multicolumn{1}{c}{\textbf{Single-Component}} & \multicolumn{1}{c}{\textbf{Multi-Component}} & \multicolumn{1}{c}{\textbf{Artifact}} & \multicolumn{1}{c}{\textbf{Task-Related}} \\ \midrule
\rowcolor{lightgray}
\textbf{p} & <0.01     & <0.01     & <0.05      & <0.01                                     \\
\textbf{t} & -3.20   & -8.19    & -2.84       & -3.17                                     \\ \bottomrule
\end{tabular}
}
\caption{T-test comparison of task completion rates within subjects. All values are significant.}
\label{tab:ttest_time}
\vspace{-10pt}
\end{table}



\begin{figure}[!ht]
    \includegraphics[width=\linewidth]{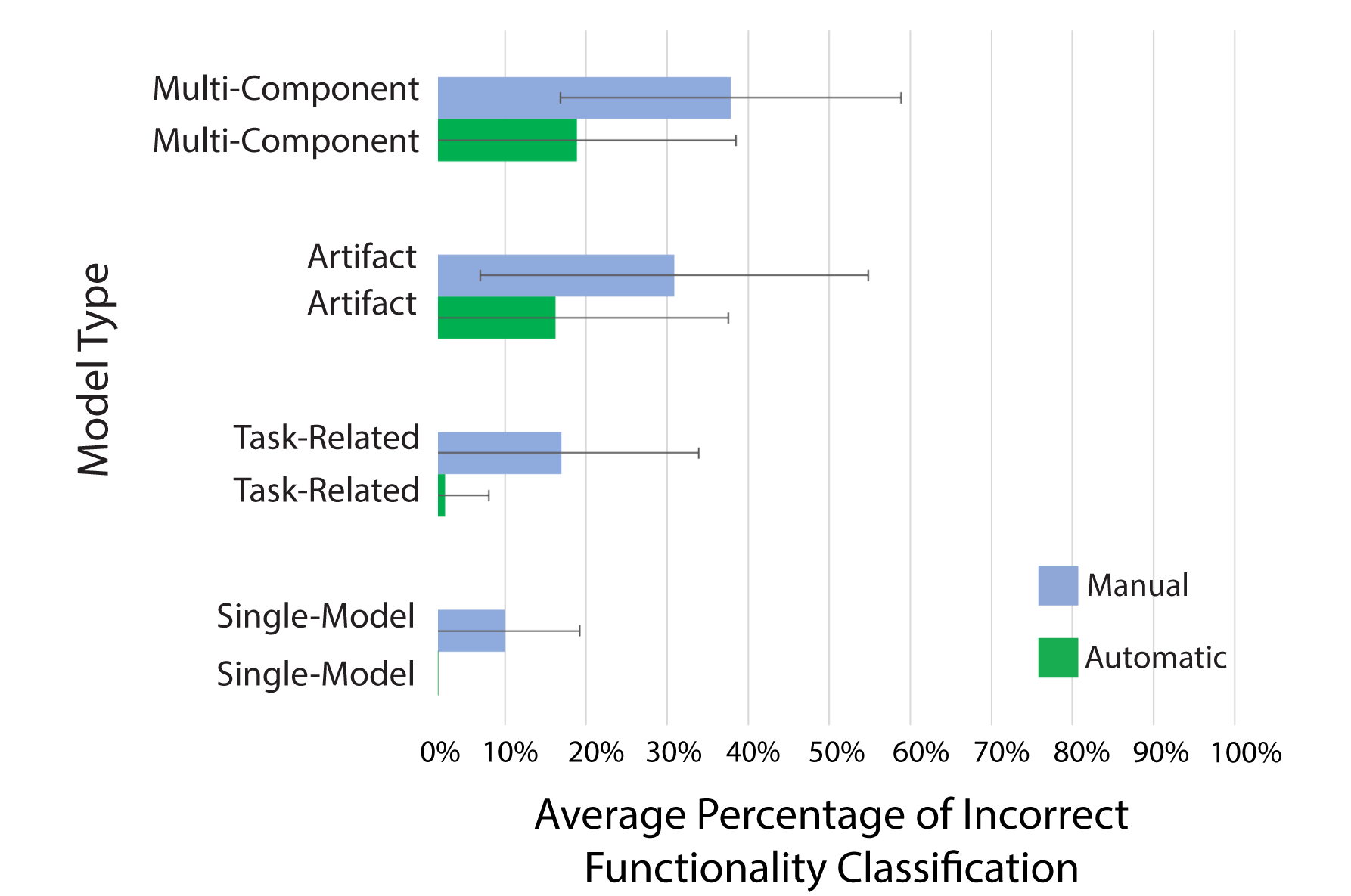}
    \vspace{-10pt}
    \caption{\changes{Box plots show the distribution of classification accuracy by condition. As the complexity of the model increased, the automatic classification helped users identify functional segments more accurately.}}
    \label{fig:user-study-acc}
\end{figure}

\changes{To analyze the functionality classification accuracy, we compared the user annotations from the study with the ground truth. To generate the ground truth, two authors annotated the models collaboratively and then fabricated the original and stylized versions to verify functionality preservation.} On comparing the user annotations with the ground truth, we found that users were more accurate in single-component models. But as the complexity increased, users had a hard time finding functional segments. The automatic classifier significantly improved performance in finding functional segments with the difference in performance increasing as the complexity increased~(\autoref{tab:ttest_acc}, \autoref{fig:user-study-acc}).

\begin{table}[ht!]
\centering

\resizebox{\linewidth}{!}{%
\begin{tabular}{@{}cllll@{}}
\toprule
 & \multicolumn{1}{c}{\textbf{Single-Component}} & \multicolumn{1}{c}{\textbf{Multi-Component}} & \multicolumn{1}{c}{\textbf{Artifact}} & \multicolumn{1}{c}{\textbf{Task-Related}} \\ \midrule
\rowcolor{lightgray}
\textbf{p} & <0.01    & <0.01       & <0.05     & <0.01   \\
\textbf{t} & -3.05   & -3.04    & -2.50        & -3.12    \\ \bottomrule
\end{tabular}%
}
\caption{T-test comparison of classification accuracy within subjects. All values are significant.}
\label{tab:ttest_acc}
\vspace{-10pt}
\end{table}

The participants generally had a positive experience stylizing models with Style2Fab. P7 said in their open-ended interview that ``\textit{being able to know if it is best as decorative or functional for areas I am uncertain of is helpful. It felt like I was approving choices rather than making them''.} P3 said \textit{``highlighting and automation made the process more precise''}, which according to P4 \textit{``makes it easier to prepare components for [...] mesh editing''}. Discussing the segmentation results, P5 also commented that \textit{``the segmentation of parts is much more uniform with Style2Fab''}, while P6 said \textit{``models were usually segmented well, but the results could be better in some more complicated models''}. P8 also said, \textit{``it usually got things right, but segmentation was too broad sometimes''}.  Commenting on the experience with the Automatic version, P6 said \textit{``this way I don't have to think through each one, I can take the suggestion and say yes or no''.} 
Overall, participants enjoyed stylizing 3D models with Style2Fab because the functionality-aware segmentation method made them confident their personalized models would work post-fabrication.

In summary, the results of the user study demonstrate significant differences between the automatic and manual conditions in terms of the time taken to process models. The analysis shows that Style2Fab's support is more beneficial as the complexity of the model increases, supporting the effectiveness of the proposed functionality-aware segmentation method.

\section{Demonstrations}

\begin{figure*}[t]
    \includegraphics[width=\textwidth]{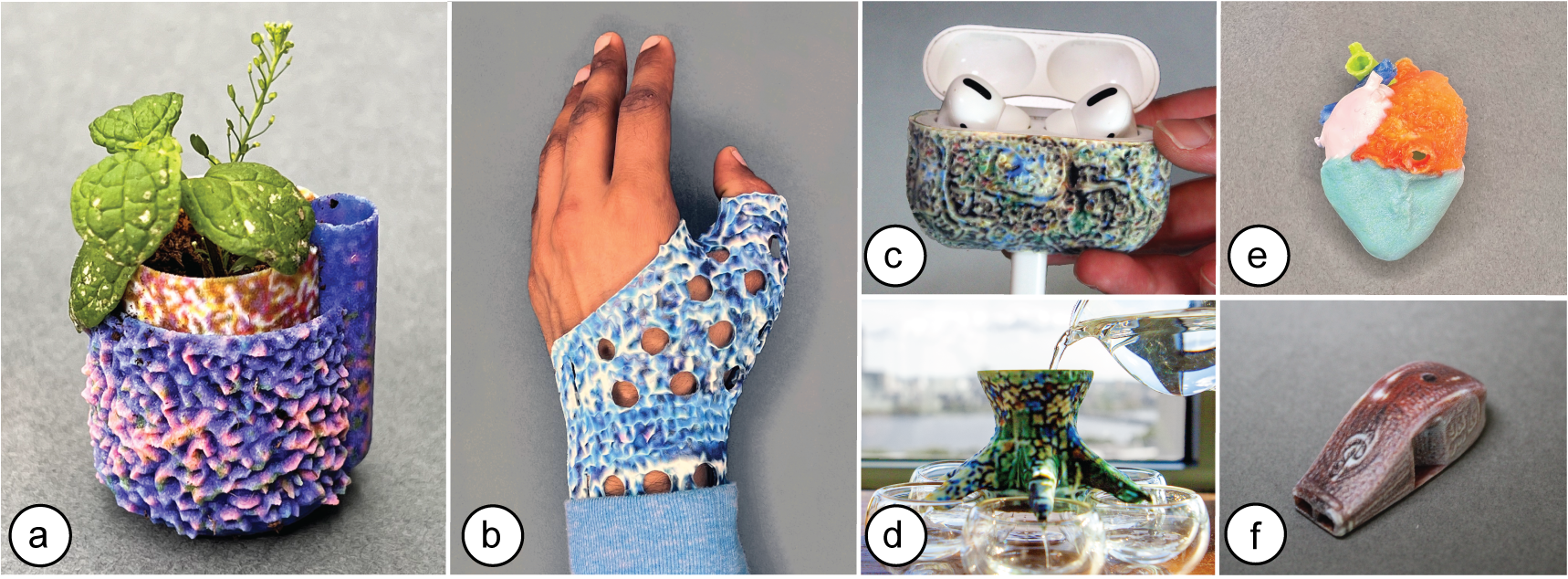}
    \caption{Application scenarios for Style2Fab (all models sourced from Thingiverse): (a) A multi-component self-watering planter styled as ``\textit{A rough multi-color Chinoiserie Planter}''. (b) A personalized Thumb Splint styled like ``\textit{a blue knitted sweater}''. (c) A personalized AirPods cover ``\textit{in the style of Moroccan Art}''. (d)  A Drinks Dispenser model styled as ``\textit{made of vintage mosaic glass tiles}''. (e) A color-coded, textured educational model of the human heart. (f) A functional whistle styled as ``\textit{A beautiful whistle made of mahogany wood}''}.
    \label{fig:application_examples}
\end{figure*}

In this section, we showcase Style2Fab's functionality-aware stylization through six application scenarios across four categories: Home Decor, Personalized Health Applications, and Personal Accessories. These are all examples of Task-Related models, and they highlight the versatility of tools that use functionality-aware segmentation to personalize models.

\subsection{Home Interior Design}
Interior design is a popular domain for personalized fabrication. Here, we demonstrate customizing a Self-Watering Planter\footnote{https://www.thingiverse.com/thing:903411/}, a Task-Related model containing two components (containing internal and external functionality). The internal functionality for this model relates to the assembly of the pot and reservoir, while externally functional segments include the base and watering cavity. Using Style2Fab, we segmented the model, verified the functional aspects, and applied the \textit{``rough multi-color Chinoiserie Planter''} style. The fabricated model showcases the desired aesthetics without compromising its self-watering capabilities or stability~(\autoref{fig:application_examples}a). Another home decor example is the Indispensable Dispenser\footnote{https://www.thingiverse.com/thing:832751/}, a drink dispenser that distributes liquid into six containers via interior cavities and spouts. We used Style2Fab to preserve the functional segments (base and interior cavities) and applied a \textit{``colorful water dispenser made of vintage mosaic glass tiles''} style to the aesthetic segments. The resulting dispenser, shown in~\autoref{fig:application_examples}d, retains its functionality while exhibiting the desired visual appearance.

\subsection{Medical/Assistive Applications}

``Medical Making''~\cite{Lakshmi_poc} and ``DIY Assistive Technology''~\cite{buehler2015sharing} are emerging and critical domains for personalized fabrication by non-technical experts. Social accessibility research~\cite{Shinohara_social} shows that considering both the aesthetic and functional features of medical/assistive devices increases their adoption. However, individuals with disabilities and their clinicians may not have the time or expertise to personalize devices~\cite{Hofmann_OT}. We first demonstrate stylizing a geometrically-complex thumb splint sourced from Thingiverse\footnote{https://www.thingiverse.com/thing:5259956} to appear as \textit{``A beautiful thumb splint styled like a blue knitted sweater.''} Like the participant from Hofmann \etal~\cite{Hofmann_OT}, we wanted the model to blend into the sleeve of a sweater~(\autoref{fig:application_examples}b). Our functionality-aware segmentation method preserved the smooth internal surface that contacts the skin and holes that increase breathability. The exterior is stylized and attractive. 

Our second example stylizes a tactile graphic of a human heart to apply unique textures to each region of the heart. This would make identifying each region easier for a blind person who accesses the model through touch. Our automatic segmentation and classification method identifies different segments of the heart that can be stylized with different textures~(\autoref{fig:application_examples}e). 

\subsection{Personalizing Accessories}
In this application scenario, we demonstrate how functionality-aware segmentation can be used to personalize accessories, such as an Airpods cover and a whistle. We demonstrate Style2Fab's personalization capabilities using a Thingiverse Airpods case\footnote{https://www.thingiverse.com/thing:4105467/files}. The interface segmented and preserved the functional aspects, including internal geometry and charging cable hole even though our classification method has no specific information about the external objects the cover interacts with. We stylized this model with the prompt: ``\textit{A beautiful antique AirPods cover in the style of Moroccan Art}~(\autoref{fig:application_examples}e). The resulting model fits the AirPods Pro case and allowed charging while featuring Moroccan Art-inspired patterns. Next, we applied styles to the popular V29 Whistle\footnote{https://www.thingiverse.com/thing:1179160} from Thingiverse without compromising its acoustic functionality. The system preserved the whistle's resonant chamber and mouthpiece while styling the exterior with a prompt: \textit{``A beautiful whistle made of mahogany wood''}. The functionality-aware styled whistle sounds like the original whistle~(\autoref{fig:application_examples}f) while the globally-styled whistle lost its functionality due to internal geometry manipulation.

\section{Discussion}
Our functionality-aware segmentation method differs from prior work because it assumes that the user will struggle to translate their understanding of the functionality of a model into key parameters of a segmentation method and specific labels for each segment. We return to the scenario of Alex stylizing her self-watering planter. Alex can recognize the pieces of the model that contribute to the self-watering functionality and the pieces that she wants to stylize. \changes{ But translating this into parameters of a segmentation method is non-trivial -- Alex would either have to go through trial and error in adjusting the hyperparameters for segmentation, or tediously highlight the segments on the vertex-level. Our functionality-aware segmentation method provides a semi-automatic method for separating aesthetic and functional segments allowing Alex to stylize her model while retaining the functionality.}


\changes{We designed our functionality-aware segmentation method to be modular and adaptable to modifications to the underlying methods. For instance, Style2Fab can be augmented with a more nuanced classifier to allow functionality beyond external and internal contexts.} In the next sections, we identify limitations and opportunities to improve on the concept of functionality-aware segmentation to adapt it to more complex and diverse domains. We expect this method could be improved by broadening our definition of functionality, creating a larger dataset of labeled classes of functionality, and using more nuanced evaluations of segment similarity.  


\subsection{A Broader Definition of Functionality}
Form and function are deeply related but do not have a one-to-one relationship; many forms can perform the same task and many tasks can be achieved with multiple forms. In our approach, we defined functionality as a topology-dependent property and ignore usage context (e.g., hanging a vase vs setting it on a table). In our formative study of Thingiverse models, the annotators infer a specific context when labeling each model's relationship between form and function. This is reliable because each model has specific affordances (e.g., a flat base affords resting on a flat surface). However, makers are creative and often play with the affordances of models to use them in new ways. 

A more nuanced approach would be to classify specific affordances. Our taxonomy presents a high-level set of affordances for interacting with external and internal contexts. However, creating a wider set of affordance-specific labels presents a trade-off. Some affordances are rare, and most classification methods struggle to label rare events. This calls for a wider set of functionality labels, beyond our taxonomy of external and internal functionality. Visual affordance is a crucial problem in robotics, with new approaches and datasets bringing insights into the domain. One such example is 3DAffordanceNet~\cite{deng20213d}which contains annotated data for 23~object categories. The essential difference between the curated deep learning datasets and online sharing platforms, is the long-tail distribution of possible designs. Although the classes represented in these datasets are represented in online repositories, the open-sharing and creative platform allows users to share novel and unique ideas, which is opposite to the standardization principle of these datasets. Thus, there is a need for novel data collection and analysis methodologies that will allow us to apply deep learning methods to analyze fabrication-oriented data.  


\subsection{Limits of Topological Similarity}
In addition to limiting our classification to two types of functionality affordances, our classification approach relies on a measure of segment and model similarity that only accounts for topological features. We selected this method because it is robust, computationally efficient, and does not rely on expensive human-generated labels. However, other measures of similarity or a measure that accounts for multiple factors may have produced a more effective classifier. \changes{More refined datasets, such as semantic labels of functionality for segments, along with a larger number of models would provide a more robust and informative approach to functionality-aware segmentation. }


\subsection{Opportunities for Richer Data Sets}
We can improve our functionality-aware classification by expanding our functionality data set and applying nuanced similarity metrics. However, like any other classification domain, this requires either larger sets of unlabeled real-world data or better labels and meta-data for existing samples. Unfortunately, 3D modeling and printing domains do not currently lend themselves to creating these types of data sets. Ideally, we could apply more advanced deep-learning methods to classify functionality, however, to account for the diversity of real-world models, this requires data sets of 3D models that are orders of magnitudes larger than our current data set.

\changes{There are multiple opportunities to curate or generate functionality information for 3D models. Makers could supply better labels by supplying well-documented source files for their meshes, but this requires a dramatic shift in the practices of these communities. More models are shared every day and new sub-domains of making are emerging with additional labels (e.g., clinical reviews on the NIH 3D print exchange \cite{Mack_NIH}). Alternatively, the creation of datasets for 3D printing~\cite{faruqi2021slicehub} and the release of novel approaches to 3D model generation~\cite{jun2023shap_e} presents an opportunity to use latent representations of 3D models to generate meta-data for functionality.}


\section{Conclusion}
In this paper, we propose a new approach to functionality-aware segmentation and classification of 3D models for 3D printing that allows users to modify and stylize 3D models while preserving their functionality. \changes{ This method relies on an insight gained from a formative study of 3D models sourced from Thingiverse: that functionality can be defined by external and internal contexts. We present our segmentation and classification method and evaluate it using functionality-labeled models from our noisy data set of real-world Thingiverse models.} We evaluate the utility of functionality in the context of selective styling of 3D models by building the Style2Fab interface and evaluating it with 8 users. This work speaks more broadly to the goal of working with generative models to produce functional physical objects and empowering users to explore digital design and fabrication.

\begin{acks}
\changes{We would like to extend our sincere gratitude to the MIT-Google Program for Computing Innovation for their generous support, which made this research possible. Furthermore, we thank Varun Jampani, Yingtao Tian, Vrushank Phadnis, Yuanzhen Li, and Douglas Eck from Google for their valuable insights and feedback on this research.}
\end{acks}

\balance
\bibliographystyle{ACM-Reference-Format}
\bibliography{references}
\end{document}